\documentclass[preprint]{aastex62}

\usepackage{subfigure}
\usepackage{bm}
\newcommand{\feh} {\mbox{\rm [Fe/H]}}

\newcommand{\oh} {\mbox{\rm [O/H]}}
\newcommand{\afe} {\mbox{\rm [$\alpha$/Fe]}}

\begin{document}

\title{Radial Migration from Metallicity Gradient of Open Clusters and Outliers}

\author[0000-0003-3265-9160]{Haopeng Zhang}
\altaffiliation{CAS Key Laboratory of Optical Astronomy, National Astronomical Observatories, Beijing 100101, China}
\altaffiliation{School of Astronomy and Space Science, University of Chinese Academy of Sciences, Beijing 100049, China}

\author[0000-0002-8442-901X]{Yuqin Chen}
\altaffiliation{CAS Key Laboratory of Optical Astronomy, National Astronomical Observatories, Beijing 100101, China}
\altaffiliation{School of Astronomy and Space Science, University of Chinese Academy of Sciences, Beijing 100049, China}

\author[0000-0002-8980-945X]{Gang Zhao}
\altaffiliation{CAS Key Laboratory of Optical Astronomy, National Astronomical Observatories, Beijing 100101, China}
\altaffiliation{School of Astronomy and Space Science, University of Chinese Academy of Sciences, Beijing 100049, China}

\begin{abstract}
Radial migration is an important process in the evolution of the Galactic disk.
The metallicity gradient of open clusters and its outliers provide
an effective way to probe for this process.
In this work, we compile metallicity, age, and kinematic parameters for
225 open clusters and carry out a quantitative analysis of radial migration
via the calculated migration distances.
Based on clusters with age $< 0.5$ Gyr, we obtain the present-day metallicity 
gradient of $-0.074 \pm 0.007$ dex/kpc. Along this gradient distributes three
sequences, and clusters in the upper, the middle, and the lower groups
are found to be old outward-migrators, in-situ clusters, and 
inward-migrators, respectively. The migration distance increases with age, 
but its most effective time is probably less than $3$ Gyr.
The metallicity gradient breaks out at $R_g$ (guiding center radius) $\sim11.5$ kpc, which is
caused by the lack of young open clusters in the outer disk and 
the presence of old outward-migrators in the upper sequence.
It shows that this boundary is related to the different effects of radial 
migration between the inner and outer disks.
We also found many special open clusters in and near the outer disk of $R > 11$ kpc
and a steeper metallicity gradient from the inner
disk of $R_g < 7$ kpc, which tells a complicated evolution history of the 
Galactic disk by different effects of stellar radial migration. 
\end{abstract}


\section{INTRODUCTION}
In recent years, with the progress of large-scale surveys in astrometry, e.g. Gaia \citep{gaia2018gaia}, and spectroscopy, e.g. APOGEE \citep{majewski2017apache}, GALAH \citep{de2015galah}, Gaia-ESO \citep{gilmore2012gaia}, and LAMOST \citep{deng2012lamost,cui2012large},
a large number of stars' positions, velocities, ages, 
and chemical abundances can be obtained, providing the most reliable 
and detailed observational constraints on the theory of the formation and 
evolution of the Milky Way. In the Galactic halo, low-$\alpha$ abundance
pattern of stellar streams provides evidence on the merging history of 
the Milky Way with dwarf galaxies in the context of the $\Lambda$cold dark 
matter model \citep[e.g.][]{nissen2018high,helmi2018merger}. In the Galactic disk, the lack of age metallicity relation
indicates an important process of stellar radial migration, which is
a recent hot topic in astrophysics \citep[e.g.][]{bergemann2014gaia,minchev2018estimating}.

According to \cite{sellwood2002radial}, the interaction between the star and 
the transient spiral arm will cause the angular momentum to change, and the 
star's guiding center will also change accordingly. \cite{minchev2010new} and 
\cite{minchev2011radial} proposed a new mechanism for radial migration, 
which is caused by the nonlinear resonance overlap between the central bar
and spiral arms of the galaxy. In addition, the disturbance caused by minor 
mergers can also give rise to radial migration 
\citep[e.g.][]{quillen2009radial,bird2012radial}. Minor mergers usually play a role
in the outer disk, but they strengthen the structure of the spiral arms and bars,
thereby indirectly affecting the entire disk \citep[e.g.][]{gomez2012signatures,minchev2013chemodynamical}.

For a long time, open clusters are widely used to trace the evolution of the Galactic disk. 
Since the ages and chemical composition of open clusters can be determined with higher accuracy than field stars, they are better tracers of the variation with time of the Galactic properties, and they can be useful to study the effect of radial migration with time. However, we should recall that open clusters are more massive than single stars, and thus the effect of the interactions with spiral arms, bars, etc. might be different and less pronounced.
In this respect, \cite{anders2017red} first suggested that 
open clusters experience significant radial migration. \cite{minchev2018estimating} 
and \cite{quillen2018migration} proposed that radial migration is expected to 
flatten the radial metallicity gradient on a long enough time scale.
Using the catalog of \cite{netopil2016metallicity}, \cite{chen2020open} 
calculated the migration distances for 146 open clusters and found that
56\% of the open clusters had migrated. They measured the migration rate
$1.5 \pm 0.5$ kpc/Gyr from intermediate-age open clusters in the outer disk,
and found a fraction of clusters tend to migrate inward.

In this work, we aim to increase the number of open clusters by combining
multiple spectroscopic survey data into the catalog of \cite{netopil2016metallicity}, and investigate
how radial migration introduces scatter in the metallicity gradient and what outlier
clusters in the metallicity gradient tell us different mechanisms of radial migration.
Moreover, since \cite{netopil2016metallicity} included mainly nearby clusters, the inclusion 
of more spectroscopic data in this work will enlarge the sky coverage,
and thus provides sufficient data for studying the radial migration of the galactic 
disk, especially the outer disk.

\section{Data and Methods}
Metallicity and age are two important parameters for estimating
the birth sites of stars, which are required in studying the radial migration 
of the Galactic disk. Based on color-magnitude diagrams from high precision photometries, 
the ages of most open clusters are available in the literature.
However, only a small percentage of open clusters have
reliable metallicities in the literature due to the lack of
spectroscopic data for many clusters. \cite{netopil2016metallicity} provided a
homogeneous sample of nearby open clusters with compiled metallicities from
previous works. They used high-resolution spectroscopic, low-resolution spectroscopic, and photometric data to derive the metallicity of 172 open clusters. We integrate this sample and only use high-resolution spectroscopic data to ensure the reliability of metallicities.
For these clusters, \cite{soubiran2018open} based on Gaia DR2 provided radial velocities.
This sample is 
the same as that of \cite{chen2020open}, but we perform a more
strict check on metallicity by comparing their values to other works.

With the recent release of high-resolution spectroscopic surveys,
including APOGEE, GALAH, Gaia-ESO, more open clusters have reliable metallicities 
and can be included in the study of radial migration of the disk.
\cite{casali2019gaia} determined the metallicity and radial velocity of 
17 open clusters from Gaia-ESO data. 
\cite{donor2020open} released the metallicity and radial velocity of 128 open clusters 
in Open Cluster Chemical Abundances and Mapping (OCCAM) survey IV, 
which are based on APOGEE DR16.
\cite{spina2021galah} obtained chemical abundances of 134 open clusters and kinematic properties of
226 open clusters from GALAH$+$ and APOGEE DR16 data.
In addition, the Stellar Population Astrophysics (SPA) survey can also provide chemical and kinematic information from high-resolution spectra of 6 open clusters \citep[see][]{frasca2019stellar,d2020stellar,casali2020stellar}.
For clusters in common in different samples, we prefer to use the value 
that using more member stars to determine the metallicity to enhance the reliability of the data.
We cross-match these samples with the cluster catalog of \cite{cantat2020painting} based on Gaia DR2,
which can provide the distance, proper motion, and age of these clusters.

Finally, the LAMOST DR7 data can provide a large number of member stars' 
spectroscopic data, including metallicity and radial velocity. 
We cross-match the LAMOST DR7 data with the list of member stars of open clusters from
\cite{cantat2020painting}, and all stars with a membership probability not less than 0.7.
The method is the same as in \cite{zhong2020exploring}, who obtained a catalog of 
295 open clusters based on the LAMOST DR5, but we have stricter selections by
limit each star with the uncertainty
of $\feh < 0.15$ dex and the uncertainty of radial velocity $< 10$ $km/s$. 
If the number of cluster members is greater than three, we calculate 
the average values (and standard deviations) of the metallicity and radial velocity.
Then we cross-match this sample with the cluster catalog of \cite{cantat2020painting}
to obtain distance, proper motion, and age information. 
We use this sample as a supplement when open clusters were not included 
in high-resolution spectroscopic data.

In order to ensure the accuracy of the following calculation, we require 
the uncertainty of the clusters’ metallicities and radial velocities 
to be less than $0.15$ dex and 10 $km/s$.
As we are interested primarily in the thin disk, we only use the sample 
with $|Z|< 0.5$ kpc 
($Z$ is the coordinate in a Galactocentric cylindrical coordinate system) 
in the following analysis to ensure that the 
metallicity gradient can be used to estimate their birth sites. 
Because most of the metallicities of clusters come from \cite{spina2021galah}, we compare the metallicities from \cite{donor2020open}, \cite{netopil2016metallicity}, \cite{casali2019gaia}, and LAMOST DR7 with those from \cite{spina2021galah}, and derive the metallicity calibrations based on clusters in common by linear fits to the data, as shown in red solid lines of Fig.~1. We do not calibrate the metallicities from 
the SPA survey because this sample is too small.
With these calibrations, we obtain a total sample of 231 open clusters with metallicities on the same scale.

\begin{figure}
\plotone{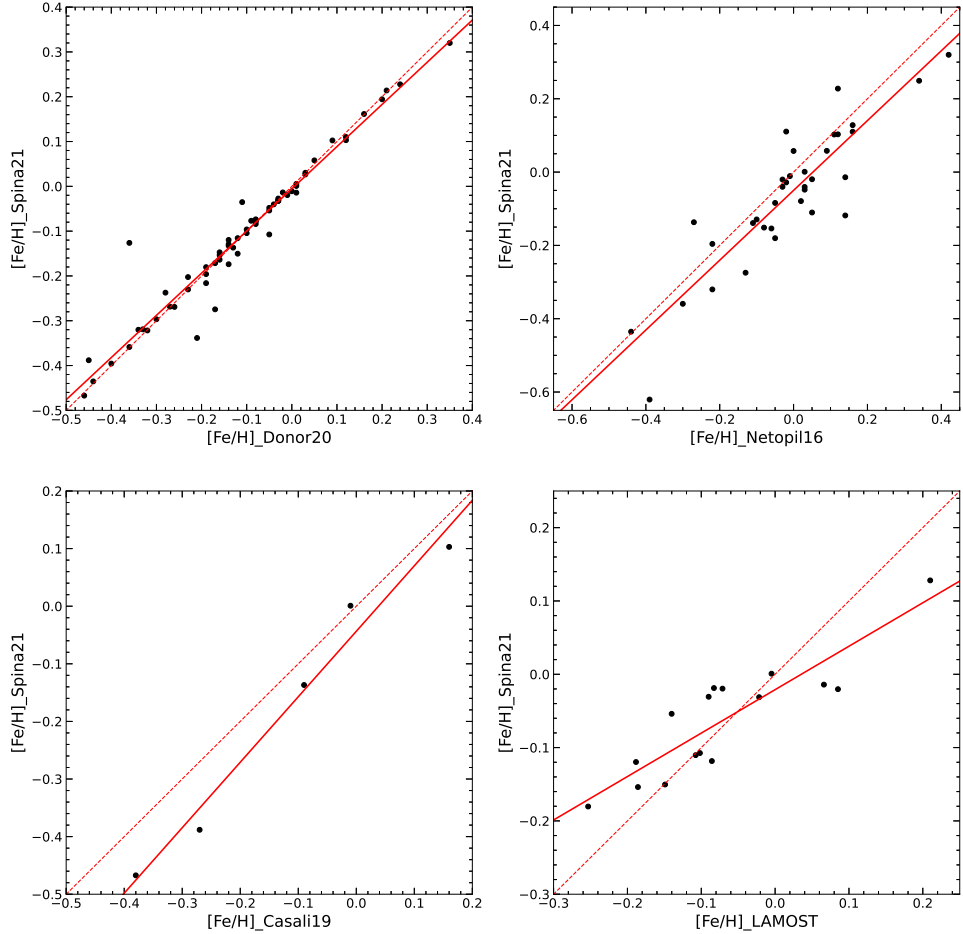}
\caption{The metallicity calibrations between \cite{spina2021galah} and
\cite{donor2020open}, \cite{netopil2016metallicity}, \cite{casali2019gaia}, and LAMOST DR7 
based on common clusters. Red solid lines are linear fits to the data and red dash lines show the one-to-one relations.
}
\end{figure}

\section{Analysis}
\subsection{Three sequences in the radial metallicity gradient}
Quantitative analysis on the effect of radial migration
requires kinematical and orbital parameters to calculate 
birth sites and migration distances.
We thus use galpy \citep{bovy2015galpy} to calculate the orbital parameters 
of the clusters. 
The gravitational potential in galpy used in our work is MWPotential2014, 
rescaled such that $R_{\odot} = 8.178$ kpc \citep{abuter2018detection}, $Z_{\odot} = 25$ pc \citep{bennett2019vertical}, and the 
circular velocity is 229 $km/s$ \citep{eilers2019circular}. We adopt 
the solar motion relative to the local standard of rest of 
($11.1, 12.24, 7.25$) $km/s$ ($U_{\odot}, V_{\odot}, W_{\odot}$, 
\citealt{schonrich2010local}). We calculate several orbital parameters of the clusters, 
such as $R_g$ (guiding center radius) and peri-/apo-center distances. 
We use the observational uncertainties of each cluster 
to calculate the errors of $R_g$ through 1000 MC runs, and the median is $0.04$ kpc.

Since the star's guiding center is a good proxy for the current orbital distance \citep{chen2020open}, 
which is not altered by blurring, and churning leads to a change 
in the star's guiding center \citep{sellwood2002radial}, so we will use $R_{g}$ rather than R to analyze the radial metallicity profile of open clusters.
Fig.~2 shows the radial metallicity gradient for the whole sample
with an X-axis of $R_{g}$. 
There are six clusters that deviate far from the general trend, and they 
are marked by their names in the three dashed boxes. Among these clusters, 
King 12, NGC 6383, NGC 2311, and Berkeley 18 have only one member star in \cite{donor2020open} and \cite{spina2021galah}, so the value could
be uncertain.
Note that the metallicity of Koposov 63
is $-0.58$ dex in \cite{spina2021galah} based on
only one member star, while \cite{zhong2020exploring} gave 
$\feh=-0.08$ dex (determined by three members).
If we adopt the latter, it does not deviate from the general trend anymore.
The metallicity of Berkeley 32 is reliable based on 11
member stars in \cite{netopil2016metallicity} and is close to the value of $-0.29$ dex
by \cite{dias2002new}. Thus, Berkeley 32 is a special cluster and will be 
discussed in Sect.~4.

Excluding these clusters in the three
dashed boxes, we obtain a total sample of 225 open clusters,
and we perform a running average metallicity for the remaining 
clusters, using a similar approach as \cite{genovali2014fine} and 
\cite{netopil2016metallicity}, by grouping the sample into a constant number 
of 15 clusters or a maximum distance range of 1.5 kpc, whichever 
criterion is met first. The result of the running average is shown by
red solid curves in Fig.~2. 
The clusters in the dashed boxes
deviate from the running average by at least 3$\sigma$, 
which also confirms their specialty.
For comparison, we perform a linear fit to the clusters 
with $R_g < 10$ kpc, and get a gradient of $-0.071$ dex/kpc. 
In general, the two sets of data are consistent until $R_g = 11.5$ kpc,
beyond which the radial metallicity gradient begins to flatten out, and
there is a significant discrepancy at $R_g\sim13$ kpc.
The reason for this discrepancy is a hot topic for decades,
but still an unsolved question today.
We also find that the radial metallicity gradient is steeper at $R_g < 7$ kpc, which will be discussed in Sect.~3.4.
Usually, the linear fitting gradient is significantly affected
by clusters on both sides in different samples, and thus the running average
may represent a more reasonable relation between $\feh$ and $R_g$.

In this work, we avoid to investigate how the metallicity gradient
flattens out in the outer disk, since we have only six open clusters
at $R_g > 13$ kpc, and they have a wide metallicity range of $\feh$
from $-0.35$ to $-0.1$ dex. Instead, we would like to analyze the
metallicity scatter at a given $R_g$ in more detail.
Generally, the scatter in metallicity is quite similar for all $R_g$,
in the order of 0.08 dex but with a wide range of 0.4 dex, which is larger than the
uncertainty in metallicity of 0.05 dex in high-resolution spectroscopic samples (the main part of
our data). It is suggested that radial migration has significant contributions 
to the scatter in the $\feh$ versus $R_g$ diagram. According to \cite{donor2020open}, 
the metallicity at a given Galactocentric distance fits well with the model
prediction of \cite{chiappini2008chemical} (without taking into account radial migration)
for young clusters (age $< 0.8$ Gyr), but
it deviates significantly from this model prediction for older clusters. This deviation
becomes more significant for the oldest clusters (age $> 2.0$ Gyr) and
is thought to be the evidence for radial migration.

Interestingly, we notice that open clusters below the dashed line seem to form a (lower) sequence parallel to the red dash line, as shown by the green dash line, which indicates that clusters with lower metallicity at the same $R_g$ also have the same gradient as the total sample.
Accordingly, we suggest that open clusters above the red dash line may represent another (upper) sequence, as shown by another green dash line.
Note that the upper sequence is also found to be a separated one consisting
of old clusters with age $> 2$ Gyr in \citet[Fig.~13]{donor2020open} based on 
high-quality data of 71 open clusters. But they do not have clusters
in the lower sequence. We check our data in the lower sequence, and 
they have consistent metallicities in the literature.
Moreover, they do not show
more considerable uncertainties in metallicity than the upper sequence.
Thus, we suggest that the lower sequence may be real, rather than a false
feature from unreliable data. In the MCM chemo-dynamical simulation 
\citep{minchev2013chemodynamical,minchev2014chemodynamical}, data 
below the theoretical prediction
of \cite{chiappini2008chemical} exist (as well as data above) as a result of
radial migration, which supports the reality of the lower sequence in our work.

\begin{figure}
\plotone{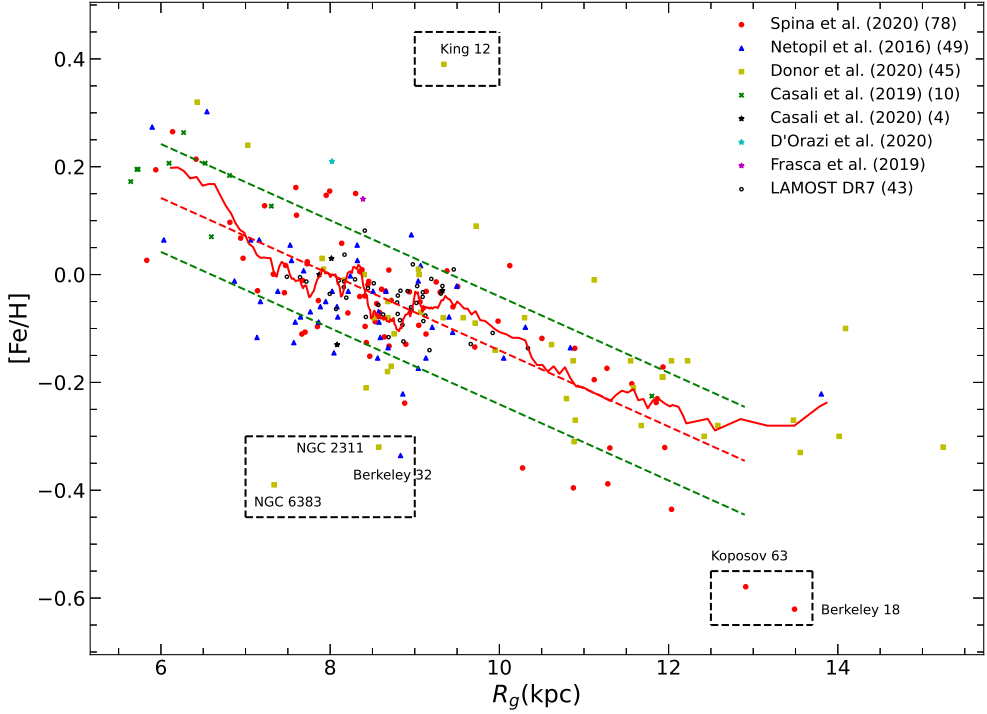}
\caption{The radial gradient of metallicity for open clusters, different symbols indicate different sources of metallicities. The clusters within the dashed box deviate from the radial gradient and are eliminated in the following work. The red solid line is the running average curve with the samples outside the dashed boxes, and the red dashed line is the linear fit to the $R_g < 10$ kpc  samples, as discussed in the text. The green dashed line is obtained by shifting the red dashed line up and down by $0.1$ dex.}
\end{figure}

\subsection{The age distributions of the three sequences.}
The age metallicity relation for the total sample is shown in Fig.~3.
The $\feh$ range of clusters is $-0.4$ to $0.3$ dex, the $R_g$ range 
of $6 - 15$ kpc, and the range of age is $0$ to $8$ Gyr. 
As expected, there is no age metallicity relation as already shown
by many previous works \citep[e.g.][]{carraro1994galactic,friel2010abundances,yong2012elemental}.
In particular, for the youngest clusters of age $< 0.5$ Gyr, there
is a $\feh$ range from $-0.3$ to $0.3$ dex, almost as large as
the whole sample. This is also the case for clusters with age $\sim 4$ Gyr.
In different R bins ($R\leq9$, $9 < R\leq12$, $R > 12$ kpc), there is also no obvious age metallicity relation, only the mean metallicity decreases from the inner to the outer disk (see Fig.~3, below).
The age metallicity relation of open clusters with age $< 4$ Gyr is similar to that of the theoretical models. For example, in the input chemical model of \cite{minchev2013chemodynamical}, the difference of $\feh$ between the present epoch and 4 Gyr ago is 0.17 dex in the solar neighborhood, which is smaller than the scatter. However, for several clusters with age $> 5$ Gyr, their metallicities are significantly higher than the theoretical values, which suggests that they may be related to radial migration.

Since radial migration requires time to take effect, young clusters
are preferred for establishing the radial metallicity gradient. In view of this,
we select the clusters with age $< 0.5$ Gyr and try to estimate the 
present-day metallicity gradient based on only
high-resolution spectroscopic data to ensure the reliability of metallicities, 
and we require at least three member stars for 
each cluster. A linear fit to the data gives $-0.074 \pm 0.007$ dex/kpc, which
is consistent with that of \cite{donor2020open} ($-0.068$ dex/kpc 
for $R_{gc} < 13.9$ kpc, $R_{gc}$ is the Galactocentric distance). The linear fit depends
significantly on the points on both sides.
Meanwhile, even for clusters with age $< 0.5$ Gyr, several open clusters
belonging to the upper and lower sequences persist, which indicates that
radial migration has a very short timescale. Therefore, it is not 
wise to limit the cluster’s age to get a reliable present-day 
metallicity gradient. Instead, we attempt to include more clusters
that do not suffer from significant radial migration in their lives.
This is actually the advantage of using the running average method, which
avoids the dependence of few points on both sides and reflects the mean
metallicity of local-born clusters (assuming that the majority of clusters
were born in local). 
Note that our derived value of $-0.074$ dex/kpc
is close to the linear fit of $-0.071$ dex/kpc in Fig.~2, and both
are consistent with the running average in the range of $R_g = 7 - 11.5$ kpc.

With the help of the derived present-day metallicity gradient,
we select a middle sequence by a shift of 0.05 dex (corresponding to the
error in metallicity)
upward and downward, and then the upper and lower clusters are
the remaining clusters in the upper and lower sequences as shown
in Fig.~5. 
The upper sequence is consistent with \cite{donor2020open}, who
suggests that the high metallicity results from their origins
of inner Galaxy and then migrated radially outwards to present
locations. On the contrary, the lower sequence indicates that
they are inward-migrated clusters from the outer disk, which
will be confirmed in terms of migration distance later.
The cumulative curves for the three sequences
in Fig.~5 indicate that the upper sequence is obviously older than the middle sequence.
As in the middle sequence, more than half of the clusters in the lower sequence are younger than 0.5 Gyr. The difference is that there are also many clusters older than 2 Gyr in the lower sequence.
Different age effects between the outward and inward
clusters may indicate different timescales or different mechanisms of radial migration.

Fig.~6 shows the distributions of clusters in the three sequences 
in the Galactic plane.
It shows that the upper sequence has the largest distribution range and 
can be distributed outside the Outer Arm.
The middle sequence corresponds to the in-situ clusters clump at the solar location near the Local Arm ($X = 8.2$
kpc, $Y = 0$ kpc), and there is no cluster near the Outer Arm.
The distribution of the lower sequence is similar to that of the middle sequence.
Again, these distributions indicate that
the effect of radial migration is related with age. The upper sequence includes
more old clusters and thus can have a large migrated coverage of radial Galactic
radius.

\begin{figure}
\centering
\subfigure{
\begin{minipage}{0.7\textwidth}
\includegraphics[width=1\textwidth]{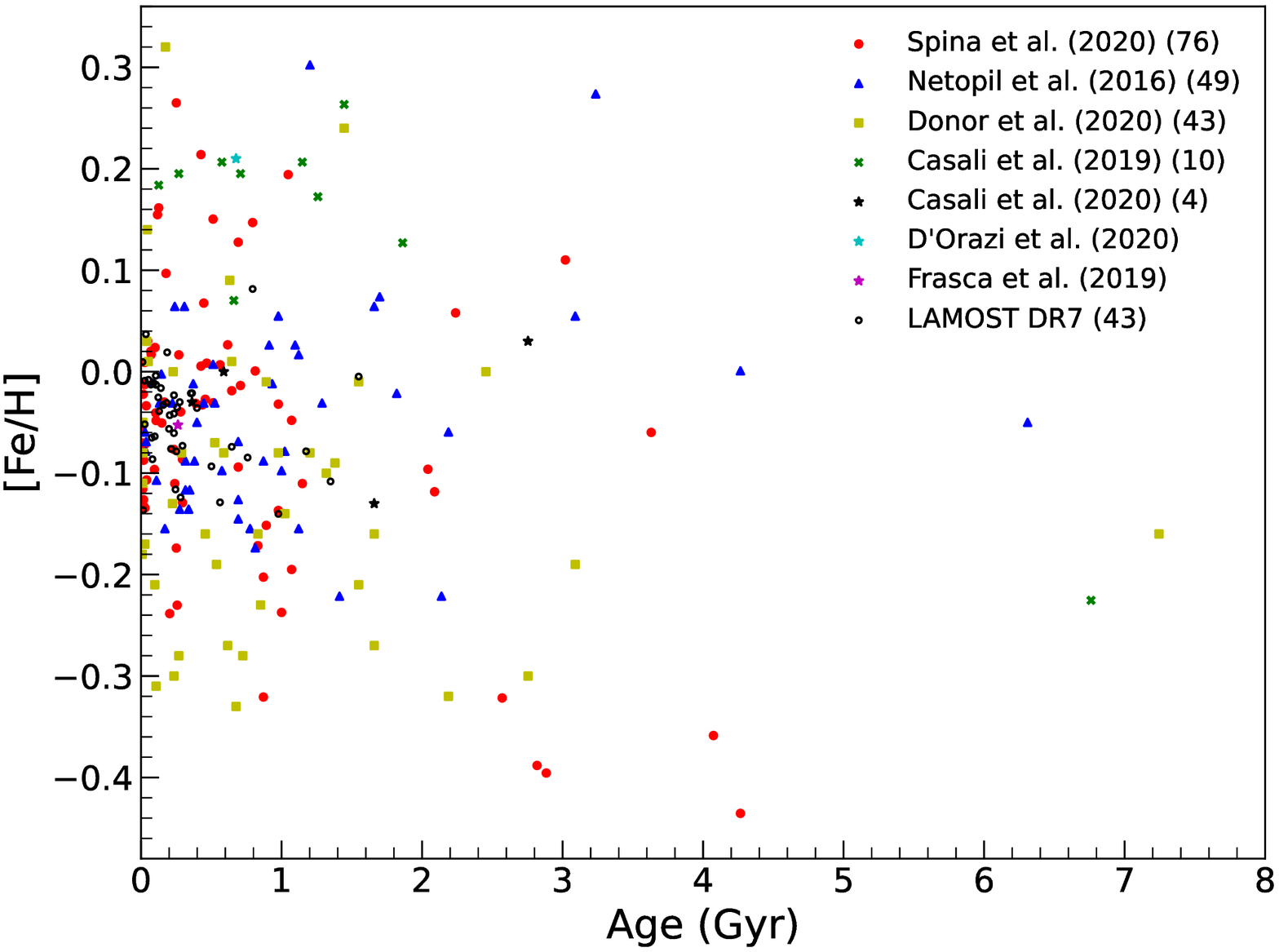}
\end{minipage}
}
\subfigure{
\begin{minipage}{0.5\textwidth}
\includegraphics[width=1\textwidth]{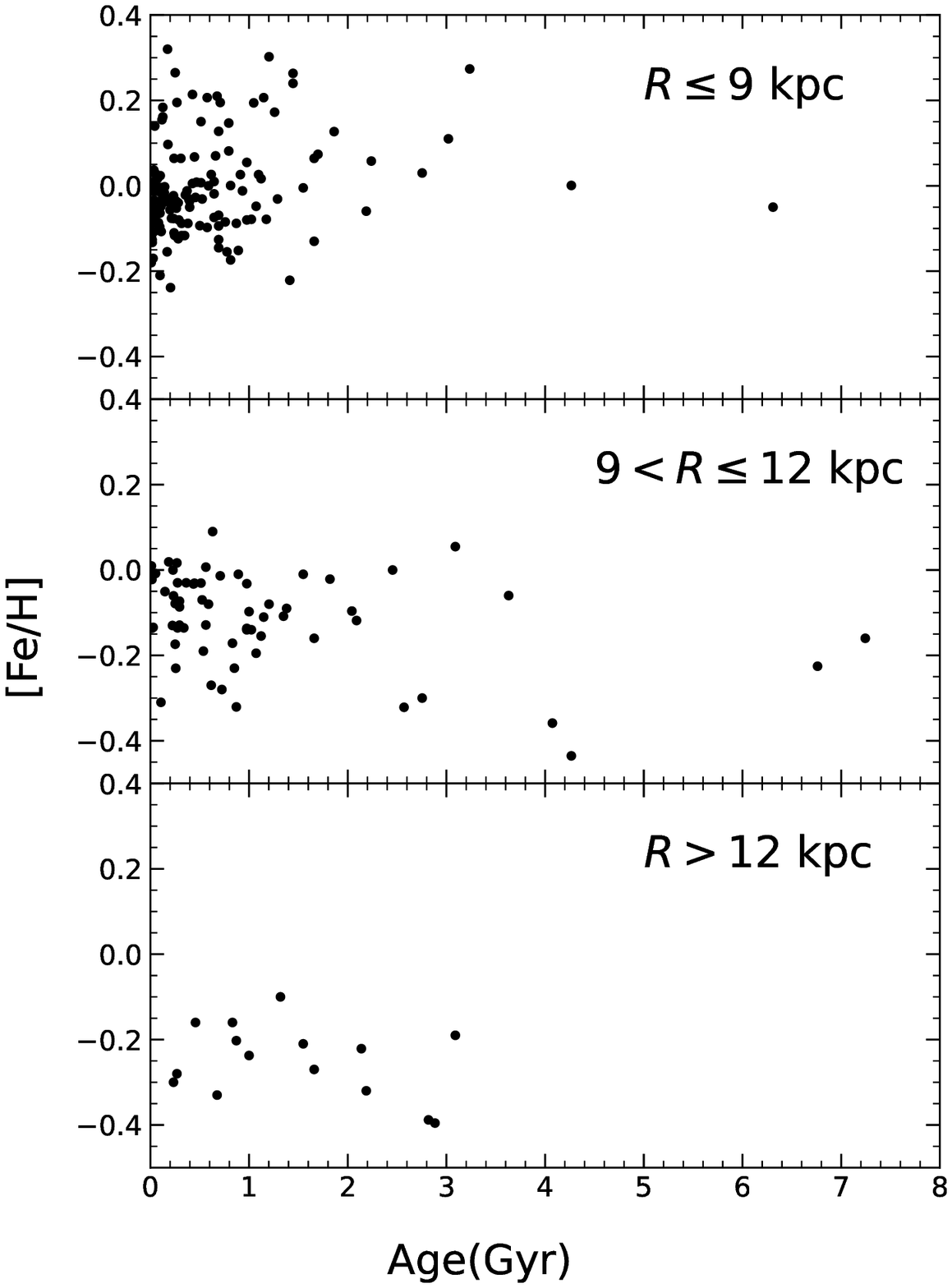}
\end{minipage}
}
\caption{
The age-metallicity relation for open clusters, the symbols are the same as in Fig.~2. The clusters within the dashed boxes in Fig.~2 have been eliminated. Age-metallicity relations for different R bins ($R \leq 9$, $9 < R \leq 12$, $R > 12$ kpc) are shown in the lower panel.} \label{fig:3}
\end{figure}


\begin{figure}
\plotone{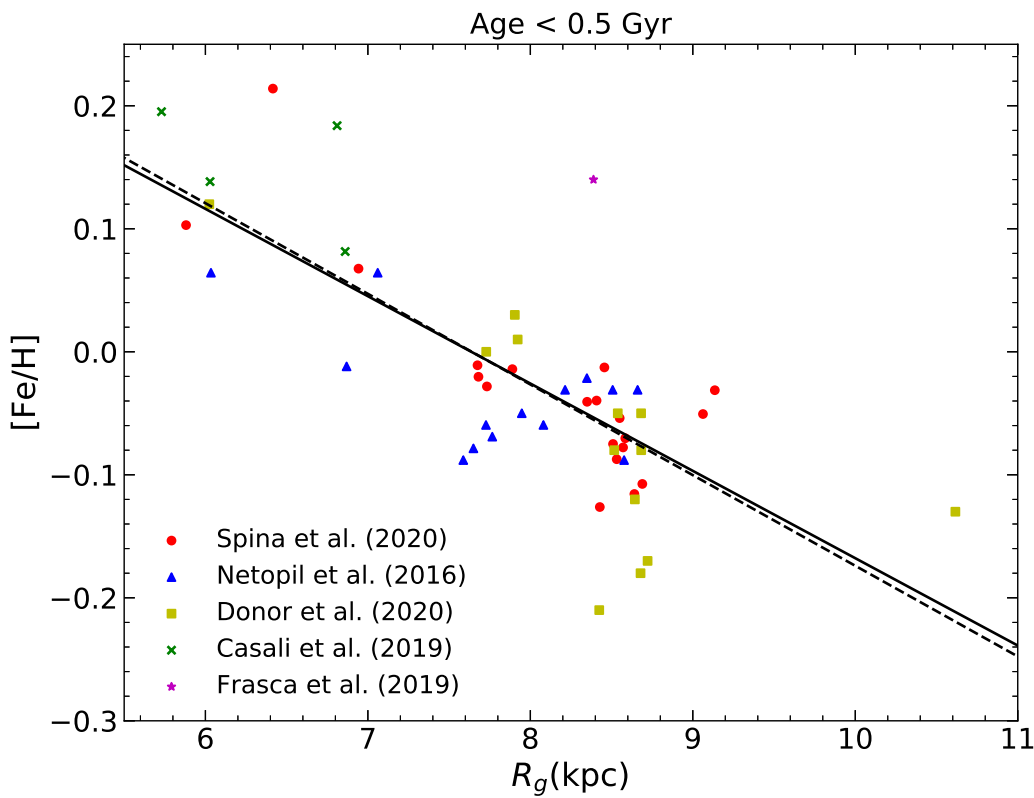}
\caption{Metallicity as a function of $R_g$ for young clusters (age $< 0.5$ Gyr) obtained by high-resolution spectroscopic data, different symbols indicate different sources of metallicities. The black dashed line is the linear fit to the data. The black solid line represents the current interstellar medium (ISM) radial metallicity distribution given by \cite{minchev2018estimating}, which has been rescaled downward by 0.14 dex.}
\end{figure}

\begin{figure}
\plotone{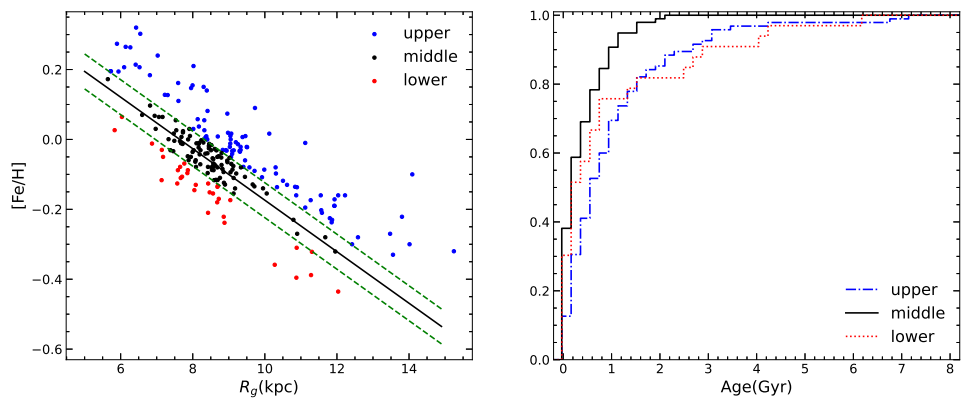}
\caption{The separation of the lower (red) and upper (blue) sequences
by selecting the middle sequence (black) along the present-day metallicity 
gradient with a shift of 0.05 dex in metallicity. The cumulative curves are shown in the right panel.}
\end{figure}

\begin{figure}
\plotone{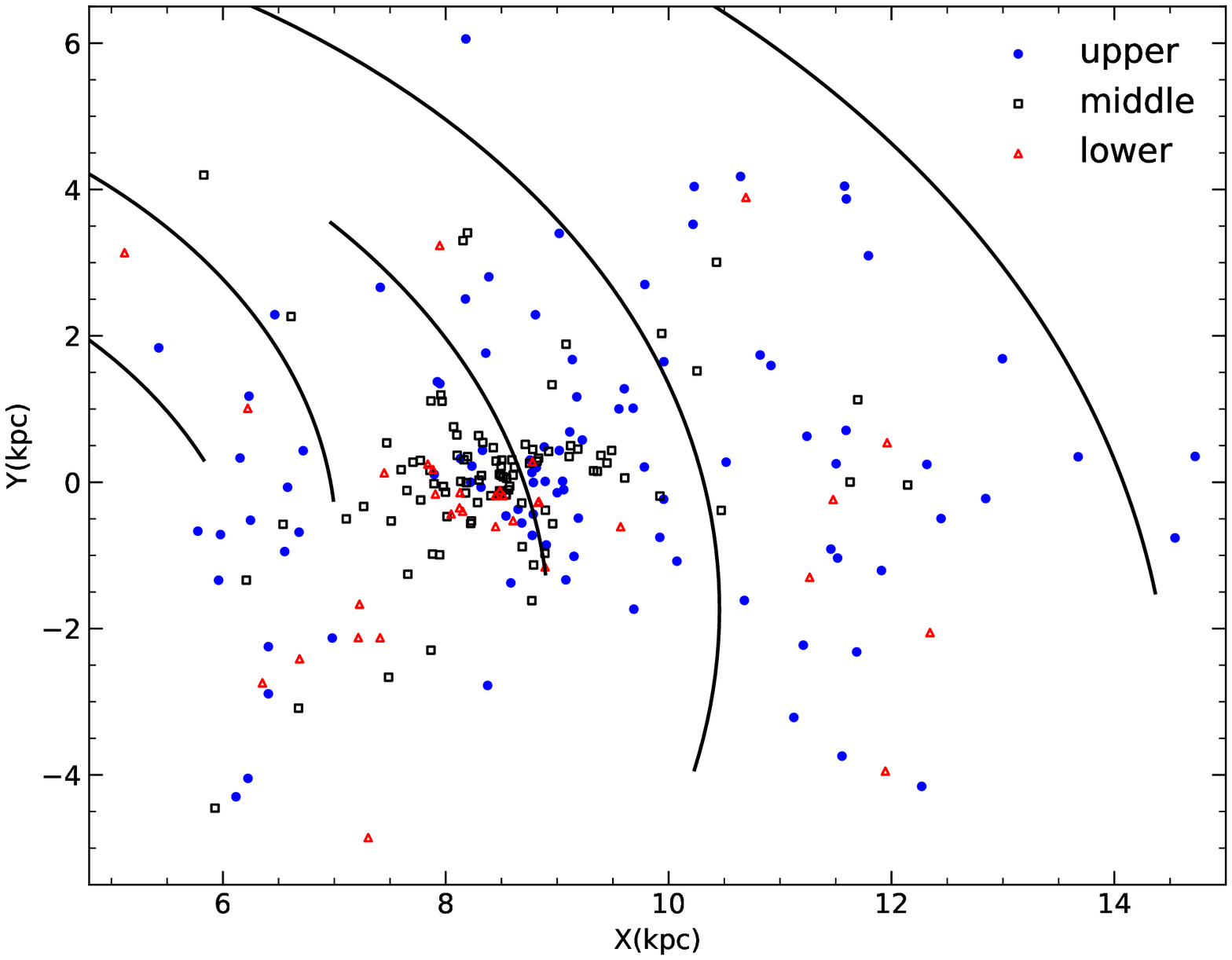}
\caption{Galactic location of the upper, middle, lower sequences projected to the Galactic plane. The Galactic center is located at (X/Y) = (0/0) kpc and the Sun at (X/Y) = (8.2/0) kpc. The black solid lines indicate the spiral arm model given by \citet[Table 2]{reid2014trigonometric}. The spiral arms from the inner to the outer area are Scutum, Sagittarius, Local, Perseus, and Outer.}
\end{figure}

\subsection{Migration distance as a function of age}

We use ISM metallicity variation with time at the solar radius and 
the ISM metallicity gradient variation with time
from \cite{minchev2018estimating} to 
calculate the birth radius ($R_b$) of clusters. 
\cite{minchev2018estimating} adopts a semi-empirical and largely model-independent method, which is based on the assumption of ISM metallicity distribution in the disk and AMBRE:HARPS and
HARPS-GTO high-quality datasets to deduce the evolution of ISM metallicity gradient with time.
This is a good alternative approach in the absence of ISM related data, because the method of \cite{minchev2018estimating} is also based on observation data.
It should be noted that an important assumption of \cite{minchev2018estimating} is that the ISM is well mixed at a given radius. The result of \cite{nieva2012present} using early B-type stars are consistent with this assumption. 
Using the HII regions data of 88 galaxies, \cite{zinchenko2016oxygen} proposed that there is no significant global azimuthal asymmetry of $\oh$ for their sample, usually lower than 0.05 dex, although some other works analyzing in the Milky Way \citep[][]{balser2015azimuthal} or individual external galaxies \citep[][]{sanchez2015census} indicate otherwise.
\cite{sanchez2015census} used NGC 6754 to conclude that the variation of oxygen abundance with azimuth is more obvious in the external Galactic regions, and the maximum scatter range of $\oh$ at a given radius is about 0.2 dex. Considering that this effect is symmetric around the mean, and most of the open clusters in our sample are in the solar neighborhood rather than in the external regions of the Milky Way, we think the assumption of \cite{minchev2018estimating} is reasonable.
However, this profile obtained 
by \cite{minchev2018estimating} using field stars may be systematically different 
from the samples of the open clusters. As shown in Fig.~4, 
for the youngest open clusters with ages of $0 - 0.5$ Gyr, according to linear fit, 
the $\feh$ of the open clusters at solar radius is -0.04 dex, while
it is $0.1$ dex in  Fig.~5 of \cite{minchev2018estimating}. 
We thus shift the ISM metallicity at the solar radius downward
by 0.14 dex in subsequent analysis. This will cause a shift
of birth site toward the inner Galaxy for the upper sequence, and
a smaller number of clusters in the lower sequence suffer
from inward radial migration.
The median error of $R_b$ is about $0.5$ kpc based on 1000 MC runs, 
using the age and metallicity uncertainties for each cluster.

In this work, the deviation of $R_{g}$ from the birth site of $R_{b}$ 
is defined to be the migration distance (MD) due to churning, and the 
deviation of present location $R$ from $R_{g}$ is due to blurring.
Fig.~7 shows the migration distance MD and $R-R_{g}$ as a function of age. 
It is obvious that churning is more significant than blurring:
about half of the clusters have $|R_{g}-R_{b}| > 1$ kpc, but only 
14 clusters lie above $|R-R_{g}|=$ 1 kpc.
Blurring has an evenly distribution around the median $R-R_{g}$ of $0.06$ kpc, 
almost close to zero. 
However, there is an increasing average of $|R - R_{g}|$ with age.
For clusters with $t \leq 0.5$, $0.5 < t \leq 1.0$, $1.0 < t \leq 2.5$, $t > 2.5$ Gyr,
the mean $|R - R_{g}|$ are 0.30, 0.38, 0.49, 0.71 kpc, respectively.
This means that blurring does not cause significant displacement
in the radial distance, but it does become more effective for old clusters.

In this work, 
we define $|MD| > 1$ kpc as a migrator, and $|MD| \leq 1$ kpc as an in-situ cluster,
based on the median error in MD of $0.5$ kpc and above 90\% clusters 
having errors less than $1$ kpc.
With this definition, 46\% of the open clusters are migrators, 
and 33\% of the youngest clusters with age $\leq 0.5$ Gyr have migrated, either
inward or outward.
In order to further analyze the change of the migration distance 
with age, we performed a running average using the same method as in Sect.~3.1. 
We adopt a maximum age range of 1 Gyr in the second criterion. The running 
average curve is shown as black solid curves in the left panel of Fig.~7. 
The average migration distance increases with age 
at age $< 3.2$ Gyr from 0 kpc to 2.9 kpc, leading to the migration
rate of approximately $1$ kpc/Gyr, 
which is similar to that estimated by \cite{quillen2018migration} ($1$ kpc/Gyr) based on the Gaussian bar model in \cite{comparetta2012stellar}, and this result is slightly smaller than that of \cite{chen2020open} ($1.5$ kpc/Gyr).
The migration rate decreases at age $> 3.2$ Gyr. We do not have enough old clusters in this analysis,
and thus, this black line stops at 4 Gyr.
It seems that the most effective time of radial migration
occurs at around the initial $3$ Gyr, with
the migration rate of $1$ kpc/Gyr.
If the results of our analysis are confirmed, this period may be related to the
time scale of radial migration caused by the interaction between the bar
and spiral arms in \cite{minchev2013chemodynamical}. 
Further work is desiring to confirm this suggestion.
In the left panel of Fig.~7, there are three clusters that deviate from other clusters. We have marked their names in the figure and will discuss them in Sect.~4 as special clusters.

We also show the MD distributions for different age bins in Fig.~8. 
For clusters with $t \leq 0.5$, $0.5 < t \leq 1.0$, $1.0 < t \leq 2.5$, 
$t > 2.5$ Gyr, the mean MD are 0.28, 0.88, 1.66, 2.00 kpc, respectively.
Again, we see the increase of the migration distance with age.
Although quite a large fraction of young clusters migrate inwards, which
reduces the mean value of MD in the first two age bins, the general
trend of increasing MD with age is valid and this is expected for radial
migration, which requires time to move from one place to another.

\begin{figure}
\plotone{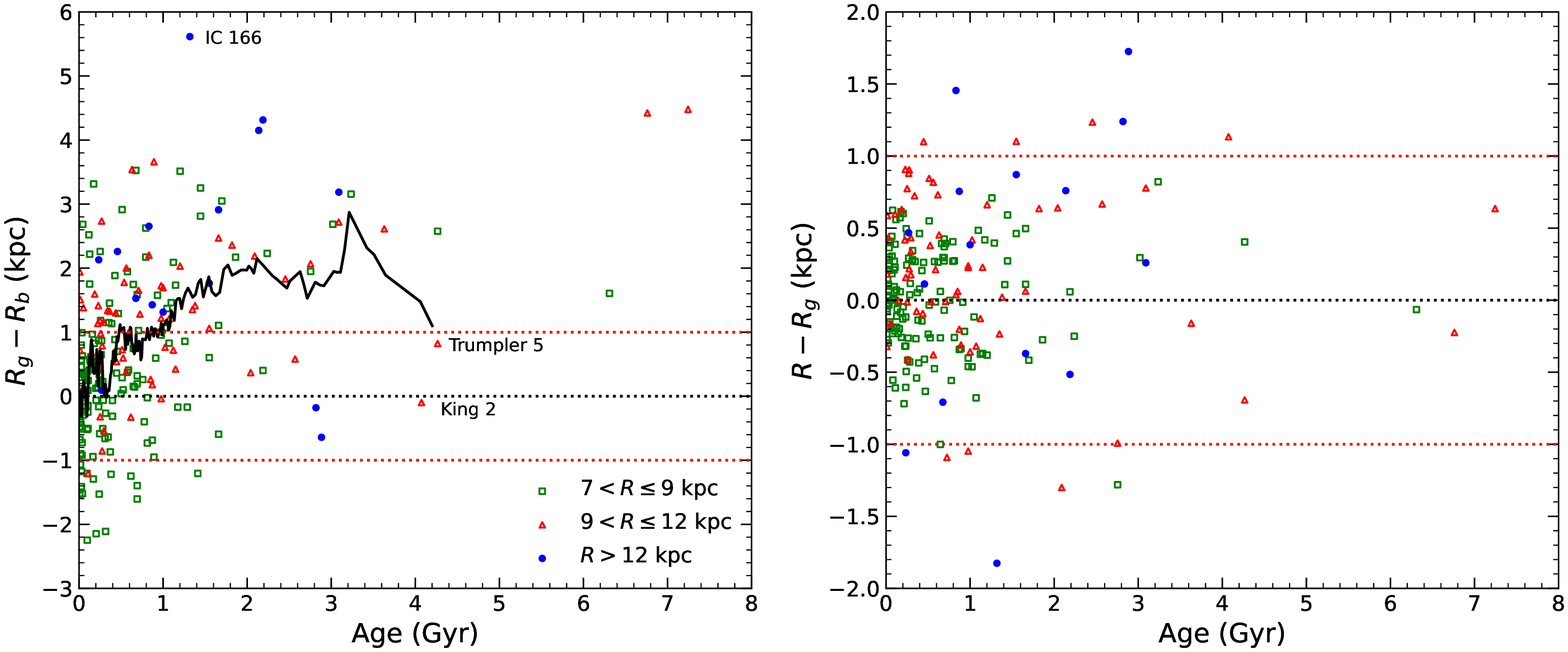}
\caption{The migration distance $R_g - R_b$ (left) and $R-R_g$ (right) versus age for different Galactic radii R bins. The black solid line in the left panel is the running average curve with all clusters.}
\end{figure}

\begin{figure}
\plotone{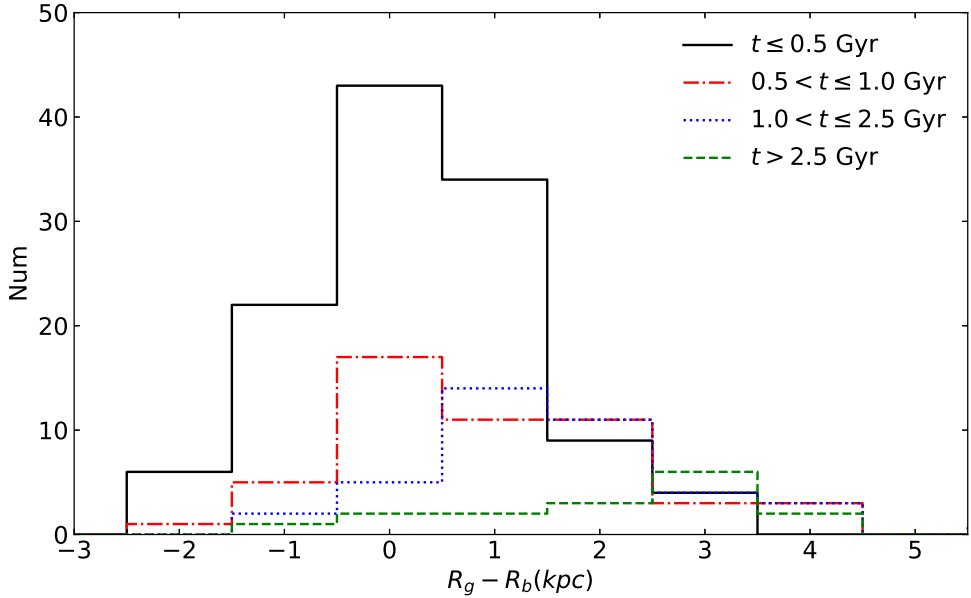}
\caption{The distributions of migration distance for different age bins: $t \leq 0.5$ Gyr (black solid line), $0.5 < t \leq 1.0$ Gyr (red chain line), $1.0 < t \leq 2.5$ Gyr (blue dotted line), $t > 2.5$ Gyr (green dashed line).}
\end{figure}

\subsection{Migration Distance as a function of $R_g$}
An obvious feature in Fig.~7 is that clusters with $R > 12$ kpc
locate the upper envelope of the MD versus age, which may be related
to the discrepancy of the metallicity gradient in Fig.~2. 
Here we adopt the $R = 11.5$ kpc as a division between the inner disk 
and the outer disk since it corresponds to the discrepancy
of the metallicity gradient between the linear fit and the 
running average in Fig.~2.
As shown in the left panel of Fig.~9, migration distances of clusters
in the inner disk and the outer disk show different trends with $R_{g}$.
The distribution of migration distance is not related to $R_{g}$ in
the inner disk where half of the clusters are in-situ, and the majority 
of the remaining clusters migrate outward with only 12 clusters 
(with young age of $t\leq 0.5$ Gyr) showing inward migration.
In the outer disk, the migration distance increases with $R_{g}$, 
and only four clusters 
are born locally. The right panel of Fig.~9 shows $R_{b}$ as a function of $R_{g}$. 
It can be seen that the birth sites of the inner and outer disk clusters are different: 
the $R_{b}$ of the inner disk clusters are mainly distributed at $3 - 12$ kpc, and 
those of the outer disk are mainly distributed at $8 - 12$ kpc. 
There are no clusters born outside of $12.1$ kpc in this sample, and most of 
the clusters in the outer disk have migrated outward to present locations.
For clusters with $R_{g} < 11.5$ kpc, $R_{b}$ increases with $R_{g}$; this trend
breaks when $R_{g}\sim11.5$ kpc, $R_{b}$ no longer increases with $R_{g}$ but flattens.
This flattening trend in the outer disk mainly occurs 
in clusters with age $> 1$ Gyr. 
In addition, all outer disk clusters
are born outside of $7.4$ kpc, and only Berkeley 36 was born within the solar radius.
Only two old clusters (Berkeley 17 and Berkeley 36) are born within the solar 
radius and have a migration distance greater than $4$ kpc. 
Their $R_{g}$ are 10.9 and 11.8 kpc respectively 
and they are all old open clusters (7.2 and 6.8 Gyr). The above results indicate the breaking
metallicity gradient is caused by the existence of old open clusters ($> 1$ Gyr)in the
outer disk beyond 11.5 kpc, which are born from the inner side and migrates to the current location. 
But clusters born within the solar radius may need a longer time to have a chance to migrate to the outer disk.

As can be seen in Fig.~9, more than half of the clusters in the inner disk of $R_{g} < 7$ kpc migrated outward to their current positions, which is higher than those near the solar radius.
The $R_b$ and $R_g$ of these outward migrating clusters are $3 - 5$ kpc and $6 - 7$ kpc respectively.
Since the clusters 
with $R_{g} < 7$ kpc are closer to the central bar, they will be more 
strongly affected by the coupling of the spiral arm and bar, and thus 
the effect of radial migration will be more obvious.
According to \cite{bovy2019life} and \cite{queiroz2020milky},
the highest metallicity is
not located in the Galactic center, but roughly at $R = 3 - 5$ kpc.
According to \cite{minchev2013chemodynamical}, the co-rotation radius is 4.7 kpc, where radial migration
is the most intense and produces inward migration as well as outward migration. 
Therefore, many metal-rich clusters migrated to $R_g > 5$ kpc, which also led to the steepening of the metallicity gradient of the clusters with $R_g < 7$ kpc in Fig.~2.

\begin{figure}
\plotone{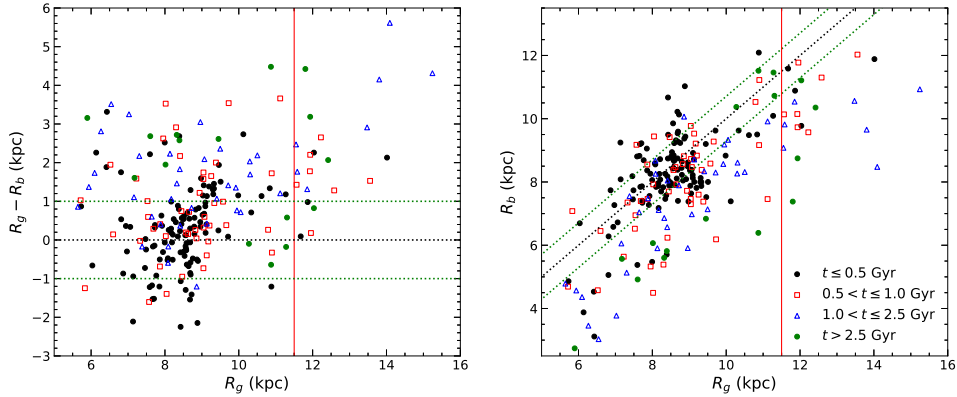}
\caption{Left: The migration distance $R_{g} - R_{b}$ versus $R_{g}$ for different age bins. The black dotted line indicates $R_{g} - R_{b} = 0$, the green dotted line indicates $R_{g} - R_{b} = 1$ or $-1$ kpc, and the red solid line indicates the boundary between the inner and outer disks, $R_{g} = 11.5$ kpc. Right: The $R_{b}$ versus $R_{g}$ for different age bins.}
\end{figure}

\subsection{MDF for migrators and in-situ clusters}
It is interesting to investigate the metallicity distribution function (MDF)
between migrators and in-situ clusters for different Galactic distances.
Fig.~10 shows the MDF of 
the total sample (upper left), $R \leq 9$ (upper right), 
$9 < R \leq 12$ (lower left), $R > 12$ kpc (lower right) samples. 
As can be seen from the upper left panel, the metallicity range of the total sample 
is $-0.45 < \feh < 0.3$ dex, which is similar to the range of migrators and in-situ clusters.
But clusters with super solar metallicity are mainly migrators, which 
form at the inner disk and migrate outward to their current locations.
Due to the high star formation rate (SFR) and high metallicity of ISM in the inner disk, we would 
expect the difference
between migrators and in-situ clusters is more significant 
at the metal-rich end,
which is also shown in the upper left panel of Fig.~10.

The MDF of in-situ clusters is picked at -0.05 dex, and has a long 
tail toward lower metallicities, indicating an obvious skewness. 
The MDF peak of the total sample 
is also at $-0.05$ dex, but the skewness is not obvious. 
The MDF of migrators has multiply peaks and tends to extend significantly toward super solar metallicity.
The skewness of the total sample, migrators, 
and in-situ clusters are $0.1$, $0.3$, and $-1.2$, respectively.
The skewness of the MDF of the total sample 
toward the low metallicities disappears due to the contribution 
of the less skewness distributed migrators, and the peak are 
determined by in-situ clusters.

We further divide the sample into three groups according to Galactic radial
distance R, $R \leq 9$ kpc, $9 < R \leq 12$ kpc, and $R > 12$ kpc.
For the largest group of $R \leq 9$ kpc, the MDF peak of the total sample is still determined 
by in-situ clusters, and migrators were evenly distributed at $-0.1 - 0.2$ dex.
The inward migrators mainly distribute in $-0.25 - -0.05$ dex, while the outward migrators mainly distribute in $-0.05 - 0.3$ dex, and there are few in-situ clusters where $\feh > 0.05$ dex.
For clusters with $9 < R \leq 12$ kpc, the characteristics 
of MDF are basically the same as those with $R \leq 9$ kpc: 
peak value of the overall sample is determined by in-situ clusters, and 
the high-metallicity tail is determined by migrators. 
The difference is that the number of inward migrating clusters is very small, which has little effect on MDF.
For clusters 
with $R > 12$ kpc, almost all clusters are outward migrators.

\cite{loebman2016imprints} used a high-resolution N-body + SPH simulation and APOGEE DR12 data to study the effect of radial migration on the MDF of field stars. They proposed that the radial migration caused the MDF to be negatively skewed at small $R_{gc}$ and positively skewed at large $R_{gc}$. Due to the small number of open clusters in our sample, we cannot analyze the skewness of MDF in each R bin to compare it with the MDF of field stars. But the same result is that their peaks are all determined by in-situ clusters, and the high-metallicity tail is almost all contributed by migrators (see their Figure 4).

\begin{figure}
\plotone{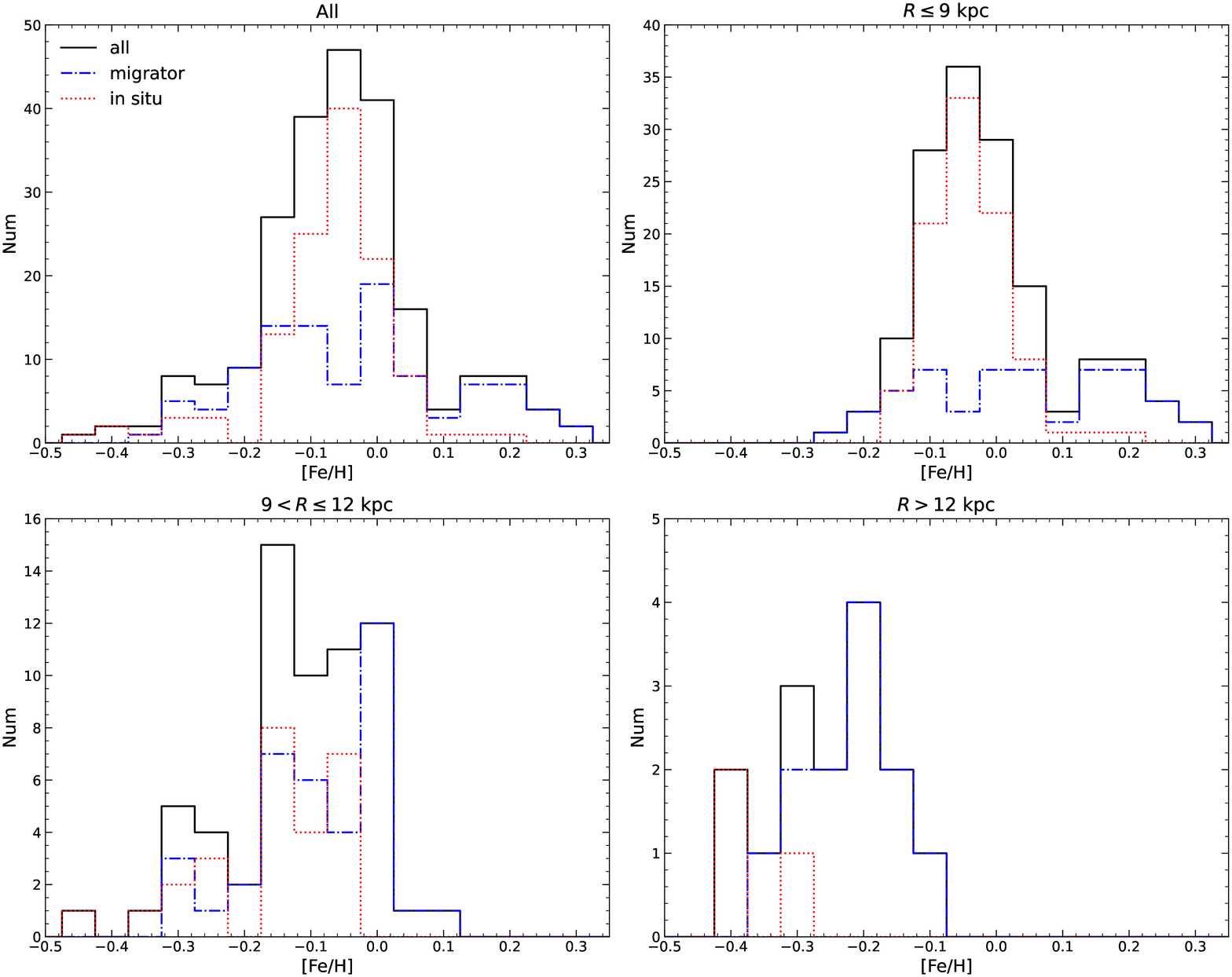}
\caption{Metallicity distribution of the total sample (upper left), $R \leq 9$ (upper right), $9 < R \leq 12$ (lower left), $R > 12$ kpc (lower right) samples. In each panel, all clusters, migrators, and in-situ clusters are represented by black solid line, blue chain line, and red dotted line.}
\end{figure}


\subsection{Migration distance for the three sequences}
It is found the metallicity of intermediate-age open clusters is higher than 
that of young open clusters at a given Galactocentric distance 
\citep[e.g.][]{jacobson2016gaia,netopil2016metallicity,spina2017gaia},
which correspond to the upper sequence in Fig.~5.
This sequence is composed of clusters that migrate outward as shown
in Fig.~11, and has an extensive $R_g$ range from the inner disk 
(5.5 kpc) to the outer disk (up to 14.5 kpc), i.e. the whole Galactic disk.
The lower sequence is mainly composed of inwardly migrated clusters, and
the $R_g$ distribution has a cutoff at 12.5 kpc, which causes 
the radial metallicity gradient to break out.
That is, there is a lack of inwardly migrated clusters 
in the outer disk (see Fig.~9), while the outwardly migrated clusters
from more metal-rich inner disk can reach farther, increasing
the mean metallicity of the outer disk and thus the
radial metallicity gradient flattening out.
In addition, radial migration leads to dispersion in radial metallicity   
distribution of open clusters, as suggested by \cite{anders2017red}.
As more clusters are migrating outward than those migrating inward,
the number of clusters in the inner disk reduces. 
They also proposed that non- and inward-migrating
clusters are disrupted faster, which can explain the present-day radial 
metallicity distribution.
According to the inside-out formation mechanism, the number density of stars
in the inner disk is higher, and open clusters are more easily to be disrupted
due to collisions. As can be seen in Fig.~2,
there are few open clusters with $R_g < 6$ kpc in our samples.

The above analysis shows different effect of radial migration between the inner 
disk and outer disk with a division around $\sim11.5$ kpc.
According to \cite{minchev2013chemodynamical},
the coupled interaction between the Galactic bar and spiral
arms could invoke very effective radial migration. This
mechanism play an important role in the inner disk, since
the Galactic bar locates in the inner Galaxy of $R < 5$ kpc
\citep[e.g.][]{bovy2019life}. For the outer disk far away from the
bar, the bar-spiral-arm mechanism may still work, but 
the impact is less as compared with the inner disk.
Meanwhile, there may be other mechanisms at work in the outer disk, 
such as minor merger \citep{quillen2009radial}, which will cause 
many special clusters to appear in the outer disk. But our sample is limited to $|Z| < 0.5$ kpc, so these special clusters are basically excluded.

\begin{figure}
        \includegraphics[width=\textwidth]{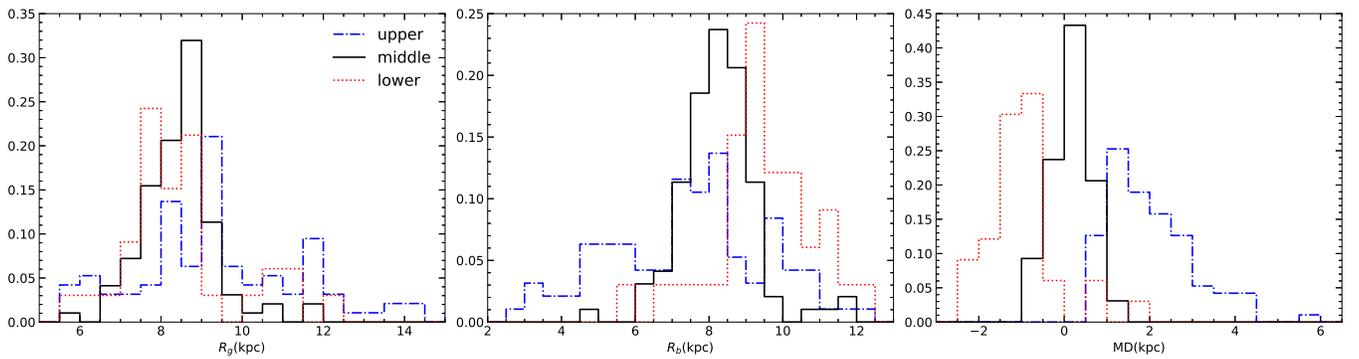}
\caption{The distribution of $R_g$ (left), $R_b$ (middle), and MD (right) of the three sequences. In each panel, upper, middle, and lower sequences are represented by blue chain line, black solid line, and red dotted line.}
    \label{fig:f11}
\end{figure}

\section{Special open clusters throughout the Galactic disk}

As discussed in Sect.~3.1, Berkeley 32 is a special cluster with underabundant
metallicity for its location.
Specifically, Berkeley 32 has a metallicity of $-0.34$ dex, 
and its age is $4.9$ Gyr. It is located at $R = 11.0$ kpc,
close to the boundary between the inner and outer disks.
Its birth radius is $9.6$ kpc and its guiding center radius 
is $R_{g} = 8.8$ kpc, and thus it is an in-situ cluster. 
However, its radial oscillation is very large, which is $4.9$ kpc.
Based on these data, it is blurring, rather than churning, that affects Berkeley 32. 

We also found three special clusters in the left panel of 
Fig.~7: King 2, Trumpler 5, and IC 166. 
The same as Berkeley 32, Trumpler 5 and King 2 are a few in-situ clusters
older than 4 Gyr with $R_{g} - R_{b}$ of $0.8$ and $-0.1$ kpc, respectively.
Therefore, there are still a few clusters
that have not to leave their birth location after $4$ Gyr.
King 2 has a similar situation to Berkeley 32: it has $\feh = -0.36$, 
age $= 4.1$ Gyr, $R = 11.4$ kpc, $R_{g} = 10.3$ kpc, $R_{b} = 10.4$ kpc.
It also has a large radial oscillation of 3.8 kpc, so the effect of blurring is larger than that of churning for King 2.
Trumper 5 has a different situation: either churning or blurring has little effect on it.
It has typical thin disk kinematics, $e = 0.08$, $Z_{max} = 0.1$ kpc, which leads to radial oscillation of only 1.9 kpc.
\cite{donati2015old} analyzed the abundances of several elements of Trumpler 5 
using high-resolution UVES spectra, and compared them with the abundance ratios of 
field stars with the same metallicity, and concluded that Trumpler 5 has a 
typical thin-disk abundance ratio (see their Figure 10). Trumpler 5 has an age of $4.3$ Gyr and
locates at $R = 11.3$ kpc.
\cite{buck2020origin} proposed that the merger brought fresh metal-poor gas
to dilute ISM's metallicity in the galactic outskirts starting from about 7 Gyr ago,
resulting in the formation of a large number of low-$\alpha$ metal-poor stars. 
According to the simulation of \cite{buck2020origin}, the age, location,
$\feh$, and $\afe$ of Trumpler 5 indicates that
it may be born after the merger of metal-poor gas.



IC 166 is currently in the outer disk, $R = 12.3$ kpc, $R_g = 14.1$ kpc, but its metallicity 
is relatively high, is $-0.1$ dex. Hence it becomes the 
most metal-rich cluster in the outer disk in the sample. 
With 1.3 Gyr age, $R_b = 8.5$ kpc, it has a high migration 
rate of $4.3$ kpc/Gyr. Also, it shows a large radial oscillation ($4.6$ kpc)
due to blurring. In short, this is a metal-rich cluster in the outer disk
with strong effect from both churning and blurring due to unknown mechanism.
Using APOGEE data, \cite{schiappacasse2018chemical} calculated 
the eight chemical abundance species (Mg, Ca, Ti, Si, Al, K, Fe, and Mn) 
for IC 166, and suggested that the cluster lies in the low-$\alpha$ sequence 
of the canonical thin disk.

In summary, for some old (age $> 1$ Gyr) 
open clusters located at $R > 11$ kpc, blurring also plays an important role, 
such as Berkeley 32, King 2, and IC 166. In addition, there are clusters that either churning or blurring has little effect on it, such as Trumpler 5, which may be born after the merger of metal-poor gas in the galactic outskirts. 
These special clusters reveal the complicated history of the evolution of 
the outer Galactic disk.

\section{conclusions}
Metallicity, age, and kinematics for a sample of 225 open clusters are compiled
from the literature to study the radial migration of the thin disk of the Galaxy. 
The metallicity and radial velocity data come from the catalog of open 
clusters of \cite{netopil2016metallicity} (only metallicity, based on a variety of high-resolution spectroscopic data), \cite{soubiran2018open}, (only radial velocity, based on Gaia DR2), \cite{casali2019gaia} (based on Gaia-ESO), \cite{donor2020open} (based on APOGEE DR16), \cite{spina2021galah} (based on GALAH$+$ and APOGEE DR16) and SPA survey, supplemented by the LAMOST DR7 data. The age data and the rest of the kinematics parameters are from \cite{cantat2020painting} based on Gaia DR2.

Based on the high-resolution spectroscopic data of clusters with age $< 0.5$ Gyr, we calculate the present-day metallicity gradient by linear fit to open clusters in the solar neighborhood, and it is $-0.074 \pm 0.007$ dex/kpc.
The systematic difference between the radial metallicity profile of 
ISM in \cite{minchev2018estimating} and our sample is modified by the present-day metallicity gradient, and the birth radius is calculated as well as the migration distance $R_g-R_b$. According to the criterion of $|R_{g} - R_{b}| > 1$ kpc, 46\% of open clusters 
have migrated.
We divide clusters into three sequences in the $\feh$ versus \b,{$R_g$} diagram according to their distribution along the present-day metallicity gradient: upper (above the fitting line), middle (near the fitting line), and lower (under the fitting line). These three sequences of clusters are mainly outward migrators, in-situ clusters, and inward migrators, respectively. 
This indicates that the metallicity scatters in the radial metallicity distribution of open clusters is mainly caused by radial migration, not only by the uncertainty in metallicity.
The $R_{g} - R_{b}$ and $|R - R_{g}|$ of clusters both increase with age, 
but the most effective time of radial migration occurs when the age is less than $3$ Gyr.
At age $> 3$ Gyr, the migration rate becomes lower.
Therefore, the radial migration of open clusters is closely related to age, and old clusters generally migrate outward.

The radial migration takes effect in an extensive range of the Galactic disk. There are many more clusters migrating outward than inward, and their distribution range is $5.5 - 14.5$ kpc, which is almost the same as the distribution range of the overall sample, while the inwardly migrating clusters are distributed in $5.5 - 12.5$ kpc. This explains why the radial metallicity distribution of open clusters has obvious scatter throughout the disk. Besides, the flattening of the metallicity gradient of the open clusters outside $11.5$ kpc can be explained by the absence of the lower sequence clusters.

The determination of the boundary between the inner and outer disks of the Milky Way is complicated, and different works give different results \citep[e.g.10 kpc in][]{haywood2013age, haywood2016milky}.
In our work, the break of the radial metallicity gradient of open clusters is found
to be 11.5 kpc where the linear function starts to deviate from the
running average curve, and the gradient flatten out. 
Furthermore, the distribution of inwardly migrating clusters has a cutoff at around $12.5$ kpc.
We thus suggested that 11.5 kpc is the boundary between the inner and outer disks based on the churning of open clusters.
We propose that the main mechanism of radial migration by the
coupling of the bar and spiral arms \citep{minchev2013chemodynamical} works in 
both the inner and outer disks, but within 11.5 kpc, this mechanism is more effective.
There are both inwardly and outwardly migrating clusters in the inner disk, but the clusters in the outer disk almost all migrate from the inner disk, and they are almost no young clusters.
Some complex extra mechanisms join the main mechanism
in the outer disk and near the boundary, which will cause some special clusters to appear, such as Berkeley 32, King 2, IC 166, and Trumpler 5.

\acknowledgments
This study is supported by the National Natural Science Foundation of China under grants No. 11988101, 11625313, 11890694, 11973048, 11927804, National Key R\&D Program of China No. 2019YFA0405502, and the 2-m Chinese Space Survey Telescope project. This work is also supported by the Astronomical Big Data Joint Research center, co-founded by the National Astronomical Observatories, Chinese Academy of Sciences and the Alibaba Cloud. 

Guoshoujing Telescope (the Large Sky Area Multi-Object Fiber Spectroscopic Telescope LAMOST) is a National Major Scientific Project built by the Chinese Academy of Sciences. Funding for the project has been provided by the National Development and Reform Commission. LAMOST is operated and managed by the National Astronomical Observatories, Chinese Academy of Sciences.

This work has made use of data from the European Space
Agency (ESA) mission Gaia (https://www.cosmos.esa.
int/gaia), processed by the Gaia Data Processing and Analysis
Consortium (DPAC, https://www.cosmos.esa.int/
web/gaia/dpac/consortium). Funding for the DPAC has
been provided by national institutions, in particular the institutions
participating in the Gaia Multilateral Agreement.


\bibliography{bibfile.bib}
\end{document}